*Article*

# Simulation Study of Coupling Effects Between a Hall Thruster and a Power Processing Unit

**Zirui Fan 1, Yinjian Zhao 1,*, Jingjing Li 1, Yingying Tian 1, Leilei Shi 2, Suliang Wu 2 and Liqiu Wei 1,**

[1] School of Energy Science and Engineering, Harbin Institute of Technology, Harbin 150001, China; [fanzirui0610@163.com] (Z.F.); [zhaoyinjian@hit.edu.cn](Y.Z.); [ljjing2025@163.com] (J.L.); [tyy15129876007@163.com] (Y.T.); [weiliqiu@hit.edu.cn] (L.W.)
[2] Shanghai Institute of Space Power-Sources, Shanghai, China; [sevenstone123@163.com](L.S.); [wusuliang_mail@163.com] (S.W.)
* Correspondence: [zhaoyinjian@hit.edu.cn] (Y.Z.)

**Abstract**

The complex and nonlinear load characteristics of Hall thrusters remain a key challenge in the design of propulsion power-supply output stages. In existing power-supply simulations for electric propulsion systems, the Hall thruster is often simplified as a fixed impedance or a prescribed current source, which makes it difficult to capture the real-time interaction between the power-supply output stage and the thruster discharge process. To address this issue, this study encapsulates a one-dimensional discharge model as an externally callable thruster slave and proposes a HallThruster.jl-Saber-Simulink co-simulation method. The proposed method enables synchronized closed-loop exchange between the power-port voltage $V_{\text{cmd}}$ and the thruster discharge current $I_{\text{out}}$ . The results show that the discharge current under the co-simulation condition exhibits a sustained low-frequency response at approximately 11 kHz. Compared with a fixed-voltage standalone simulation, the co-simulation shows observable differences in port waveforms, spectral characteristics, and internal field distributions. This method provides a co-simulation basis for realistic load analysis of propulsion power supplies and subsequent stress evaluation of key components.

## 1. Introduction

Hall thrusters have been widely used in station keeping, orbit transfer, deep-space exploration, and low-Earth-orbit satellite constellations because of their high specific impulse, long lifetime, and high propellant utilization efficiency [1]–[3]. As the on-orbit operating time and the number of start-stop cycles of electric propulsion systems continue to increase, system design has gradually shifted from the performance of an individual thruster to the coordinated matching among the thruster, the power processing unit (PPU), and the control system. As the core unit for energy conversion and control in an electric propulsion system, the PPU must provide multiple power supplies for anode discharge, magnetic excitation, cathode heating, ignition, and other subsystems [3]–[6]. In particular, the anode power-supply output stage is directly connected to the thruster discharge channel and provides the anode voltage for the thruster. It is also subjected to output ripple and device stress induced by discharge-current oscillations and transient load variations. Therefore, it is a critical part of propulsion power-supply reliability design.

In current PPU design practice, the thruster load is often represented as a fixed resistor, an equivalent impedance, or a prescribed current source. These approaches are convenient for analyzing power topology, control loops, output filters, and device stress, but they cannot represent the real-time feedback of the thruster discharge state to the power-supply output stage. In fact, a Hall thruster is not a constant-impedance load. Neutral transport, ionization, electron transport, potential reconstruction, and ion acceleration inside the thruster jointly affect the discharge current and can generate periodic low-frequency oscillations [7]–[16]. Therefore, if a fixed impedance or an offline current waveform is used as the thruster load in the analysis of a propulsion power-supply output stage, the real-time interaction between the thruster discharge process and the power port may not be captured, and the output-voltage ripple, load fluctuation, and subsequent device stress may be underestimated.

Previous Hall-thruster studies have mainly focused on discharge physics, low-frequency oscillation mechanisms, and internal plasma evolution. Choueiri systematically reviewed plasma oscillations in Hall thrusters [7]. Boeuf and Garrigues modeled the formation of low-frequency oscillations in stationary plasma thrusters [8]. Lafleur et al. analyzed the origin and stabilization of the breathing mode using 0D and 1D models [9]. Dale and Jorns studied the time-resolved variation of plasma parameters during the breathing mode [10]. Chapurin et al. discussed the formation mechanism of low-frequency instabilities from the perspectives of ionization-oscillation models and fluid/hybrid simulations [11], [12]. Petronio et al. further investigated the development of the breathing mode using hybrid simulations [13]. Leporini et al. studied the onset and development of the breathing mode from the perspectives of electron-mobility perturbations and unstable 0D ionization models [14], [15]. In contrast, PPU-related studies have mainly focused on power architecture, power density, and engineering implementation. Fang et al. proposed a PPU system architecture based on a high-frequency AC bus [5]. Goebel and Katz systematically described the power-supply relationships among the thruster, cathode, and PPU in electric propulsion systems [3]. These studies have advanced both Hall-thruster discharge modeling and PPU engineering design. However, in most simulations, the two parts are still connected through simplified boundaries or equivalent loads: the thruster side usually adopts a prescribed voltage boundary, whereas the power-supply side often treats the thruster as a fixed impedance or a prescribed current source. This treatment facilitates separate analysis of the thruster and the power supply, but it is difficult to describe the real-time interaction between the power-supply output stage and the thruster discharge process.

In recent years, the influence of power-circuit parameters on Hall-thruster discharge dynamics has attracted increasing attention. Brieda et al. showed that the power-supply model can affect simulated discharge-voltage oscillations of Hall thrusters [17]. Simmonds et al. investigated the discharge response of a Hall thruster under modulated voltage and its relationship with the breathing mode [18]. Jovel et al. carried out impedance characterization of Hall thrusters, providing a basis for understanding the dynamic thruster load from the power-supply side [19]. Jorns et al. modeled mode transitions in a magnetically shielded Hall thruster, further indicating that discharge dynamics are closely related to external boundary conditions [20]. These studies demonstrate that the PPU and the Hall thruster do not form a simple one-way power-supply relationship. The power-supply output stage can affect thruster discharge parameters, while the thruster discharge state also feeds back to the power port through dynamic current. Therefore, an online coupling method capable of synchronously exchanging port voltage and discharge current during simulation is required.

Based on these issues, this paper establishes a port-level online co-simulation model for the dynamic load characteristics of Hall thrusters using HallThruster.jl [21], Simulink [22], and Saber [23]. In this model, the open-source one-dimensional Hall thruster discharge solver is encapsulated as a physical load module that can be driven in real time by the output stage of the power supply. Through the Simulink interface, the output-port voltage from the Saber power supply model and the discharge current from HallThruster.jl are synchronously exchanged, enabling closed-loop time-domain coupling between the power supply response and the internal discharge state of the thruster. During the simulation, in addition to the port voltage, discharge current, and frequency-domain response, internal thruster parameters such as electron density, neutral density, electric potential, axial electric field, and electron temperature can also be recorded. These results can be used to examine whether variations in the power-supply port voltage affect the internal discharge process of the thruster. The proposed method provides an extensible co-simulation platform for studying the dynamic coupling between the power supply and the Hall thruster, and also provides a basis for subsequent analyses of output voltage ripple, load matching, and key component stress.

## 2. Materials and Methods

This section introduces the model components and data-exchange strategy of the online co-simulation. The Saber side is used to build the propulsion power-supply output-stage model and to describe the transient responses of the power-port voltage, current, and output filter network. The HallThruster.jl side is used to calculate the one-dimensional Hall-thruster discharge process and to obtain the internal plasma state and discharge current under the current discharge boundary. Simulink serves as the interface layer and organizes the data transfer and synchronous calls between Saber and Julia.

### *2.1 Hall-Thruster Fluid Model*

In propulsion power-supply output-stage simulation, the choice of the load model directly affects the calculated port voltage, current ripple, and component stress. For a conventional resistor, the port relationship is mainly determined by fixed parameters. For example, a resistive load satisfies the approximately linear relation $I = V/R$. Under a given operating condition, the parameters of this type of load usually remain unchanged, and the power-supply side can describe its port response using a fixed impedance or simple differential equations.

A Hall thruster is different. Although the anode port of the thruster appears as a voltage-current relationship at the circuit level, its discharge current is not determined solely by the instantaneous port voltage. Instead, it is jointly determined by the neutral density, electron density, potential distribution, axial electric field, electron temperature, and ionization source term in the discharge channel. In other words, the propulsion power supply does not see a fixed impedance element, but a nonlinear dynamic load formed by the real-time evolution of the plasma state. A change in the port voltage modifies the electric field and electron-energy state in the channel and then affects the ionization rate and charged-particle transport. Conversely, variations in particle density and potential structure change the discharge current and feed back to the power-supply output stage. Therefore, the equivalent load of a Hall thruster is clearly time-varying and nonlinear and cannot be simply treated as a fixed impedance.

This study uses the one-dimensional axial discharge model in HallThruster.jl to describe the main plasma processes inside the thruster. Along the axial direction, the model solves governing relations related to neutral transport, ion generation and acceleration, electron transport, potential distribution, and electron-energy evolution. The neutral continuity equation describes propellant supply and ionization depletion. The ion continuity equation and momentum relation describe ion generation and axial acceleration. The electron-transport relation describes electron mobility under magnetic-field confinement. The potential and axial electric field jointly determine the position of the main acceleration region. The electron-energy equation reflects the balance among electron heating, ionization loss, and wall loss. These internal variables jointly determine the discharge current at the current time.

To clarify the relationship between the port current and the internal state, the governing relations directly related to the formation of the dynamic load are listed below. Here, z denotes the axial coordinate, ne is the electron density, nn is the neutral density, Te is the electron temperature, ϕ is the potential, $E_z$ is the axial electric field, $I_d$ is the discharge current, and$S_{ion}$is the ionization source term. The specific numerical discretization, boundary conditions, and transport closures follow the original implementation of HallThruster.jl.

To avoid ambiguity in the sign convention, the axial coordinate $z$ is defined from the anode toward the cathode/exit region. The electrostatic potential is denoted by $\phi$, and the axial electric field is defined as

$$E_z = -\partial\phi/\partial z \tag{1}$$

The ion and electron particle fluxes are defined positive in the $+z$ direction. The ion flux is $\Gamma_i = n_i u_i$, where $n_i$ is the ion density and $u_i$ is the axial ion velocity. The electron particle flux is denoted by $\Gamma_e$. Under this convention, the discharge current through the channel cross-sectional area $A$ is written as

$$I_d = eA(\Gamma_i - \Gamma_e) = eA(n_i u_i - \Gamma_e) \tag{2}$$

Here, $e$ is the elementary charge. The negative sign before $\Gamma_e$ accounts for the negative charge of electrons when the electron particle flux is defined in the positive axial direction.

The neutral continuity equation describes propellant injection and ionization depletion:

$$\partial n_n/\partial t + \partial(n_n u_n)/\partial z = -S_{ion} \tag{3}$$

Where $n_n$ is the neutral density, $u_n$ is the axial neutral velocity, and $S_{ion}$ is the ionization source term. The ion continuity equation is

$$\partial n_i/\partial t + \partial(n_i u_i)/\partial z = S_{ion} \tag{4}$$

The one-dimensional ion momentum equation can be written in conservative form as

$$\partial(n_i u_i)/\partial t + \partial(n_i {u_i}^2 + p_i/m_i)/\partial z = (e/m_i) n_i E_z \tag{5}$$

where $p_i$ is the ion pressure and $m_i$ is the ion mass. If multiple ion charge states are considered, the same form is applied to each charge state with the corresponding charge number.

Electron transport is described using the quasineutral drift-diffusion approximation. The axial electron particle flux can be expressed as

$$\Gamma_e = -\mu_\perp n_e E_z - (\mu_\perp/e)\partial p_e/\partial z \tag{6}$$

where $n_e$ is the electron density, $p_e$ is the electron pressure, and $\mu_\perp$ is the effective cross-field electron mobility. The mobility is written as

$$\mu \perp = e/[meve(1 + \Omega e^2)], \Omega e = \omega ce/ve, \omega ce = e|B|/me \tag{7}$$

Here, $m_e$ is the electron mass, $B$ is the radial magnetic-field strength, $\omega_{ce}$ is the electron cyclotron frequency, and $\Omega_e$ is the Hall parameter. The total electron momentum-transfer collision frequency is

$$\nu_e = \nu_c + \nu_A N \tag{8}$$

where $\nu_c$ is the classical collision frequency and $\nu_{AN}$ is the anomalous collision frequency. For the default anomalous-transport closure used here, HallThruster.jl adopts the TwoZoneBohm model:

$$\nu_{AN} = c_1 \omega_c e, \quad z < L_{ch}; \qquad \nu_{AN} = c_2 \omega_c e, \quad z \geq L_{ch}, \tag{9}$$

where $c_1 = 1/160$ inside the channel and $c_2 = 1/16$ downstream of the channel exit unless otherwise specified in the input file. $L_{ch}$ is the channel length.

The electron-energy balance is written as

$$\partial W_e / \partial t + \partial q_e / \partial z = Q_J - Q_{ion} - Q_{wall} - Q_{coll} \tag{10}$$

where $W_e$ is the electron internal-energy density, $qe$ is the electron-energy flux, $Q_J$ is the Joule heating term, and $Q_{ion}$, $Q_{wall}$, and $Q_{coll}$ are the ionization, wall, and collision loss terms, respectively. The detailed numerical discretization, boundary conditions, anomalous transport closure, wall-loss treatment, and initial condition follow the HallThruster.jl implementation used in this work and are summarized in Table 1.

Table 1. Main parameters of the HallThruster.jl-Simulink-Saber co-simulation

| Parameter | Value or setting used in this work |
|---|---|
| Propellant | Xenon (Xe) |
| Anode mass flow rate | 5 mg/s |
| Thruster type | SPT-100 predefined thruster |
| Inner radius | 0.0345 m |
| Outer radius | 0.0500 m |
| Channel length $L_{ch}$ | 0.025 m |
| Axial domain | 0-0.08 m |
| Number of axial cells | 100 |
| Magnetic-field profile | SPT-100 Gaussian B-field (built-in) |
| Magnetic-field scale | B_scale = 1.4 for the main case |
| Anode potential boundary | Dirichlet potential boundary at the anode; the discharge voltage is provided by the sampled Saber output-port voltage $V_{cmd}$ |
| Cathode/exit potential boundary | Cathode/exit potential set to 0 V unless otherwise specified; potential equation uses Dirichlet boundaries at anode and cathode |
| Electron transport closure | Quasineutral drift-diffusion approximation / generalized Ohm law |
| Effective electron mobility | $\mu_\perp = e/[m_e \nu_e (1 + \Omega_e^2)]$, $\Omega_e = \omega_{ce}/\nu_e$ |
| Anomalous collision model | TwoZoneBohm($c_1$, $c_2$), HallThruster.jl default anomalous-transport model |
| Wall-loss model | ConstantSheathPotential(-20.0, 1.0, 1.0), unless overridden |
| Initial plasma state | Initialized internally by HallThruster.jl from the input configuration; no offline current waveform is imposed |
| HallThruster.jl internal time step | 5 ns |
| PPU/Saber port sampling interval | 1 μs in the present manuscript; |

*2.2 HallThruster.jl-Simulink-Saber Co-Simulation Method*

Figure 1 shows the port-level online co-simulation structure adopted in this work. The Saber side represents a propulsion power-supply output stage and its port transients. Simulink serves as the interface organization layer and is responsible for S-Function calls and TCP data forwarding. The Julia side runs the HallThruster.jl thruster slave. Only two port variables are exchanged among the three components: Simulink-Saber sends $V_{cmd}$ to Julia, and Julia returns $I_{out}$, which is determined by the current internal discharge state.

The key feature of this structure is the synchronized feedback of port variables. At each interface instant, the Saber output stage receives $I_{out}$ returned by the thruster model from the previous communication step. The Julia thruster slave updates $V_{cmd}$ from the power-supply side with each communication step. Therefore, the port voltage, the internal thruster state, and the discharge current returned to the power-supply side are updated sequentially within the same time-marching process, forming an online closed loop of "port-voltage update of the discharge boundary - internal plasma-state advancement - diagnostic-current feedback to the output stage."

The Saber circuit model retains the component architecture directly related to port dynamic response, so that the output voltage can be instantaneously adjusted under finite-output-impedance conditions according to the returned discharge current. With this treatment, the Saber side provides a non-ideal power-supply output, while the Julia-side thruster model updates the discharge state in real time and returns the dynamic load current.

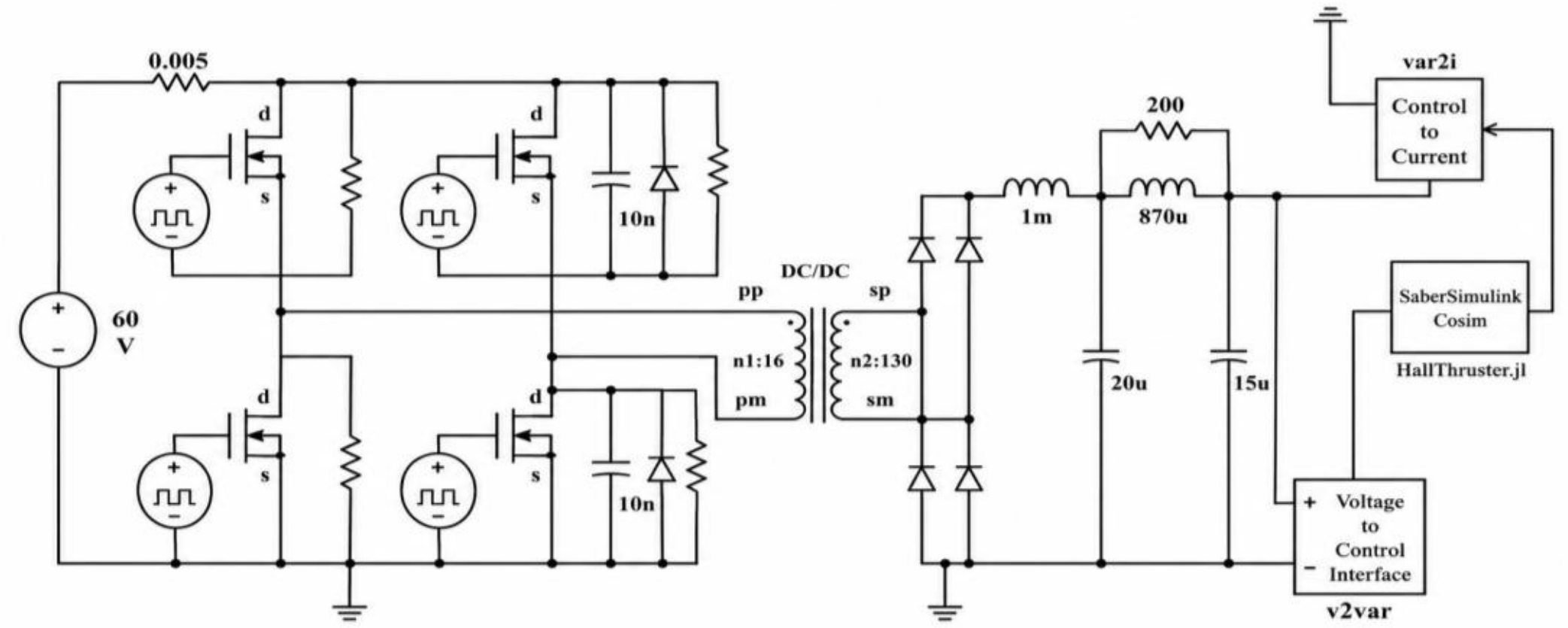

Saber–Simulink–Julia Co-Simulation Architecture

Master Clock (Saber)

Plant Response (Julia)

Saber
Circuit Solver
*Power Supply Chain*

Source · Switch · Transformer · Rectifier Filter · Load

Nodal Voltage / Simulation Step (Vcmd, At_macro)

Equivalent Current Feedback (Iout)

Simulink / S-Function

SIMULINK®

- Scheduling and Time Management
- Coupling Logic
- Data Packing / Unpacking
- TCP Client Communication

Vcmd, At_macro

Iout

Julia / HallThruster.jl

HallThruster.jl

Figure 1. Port-level co-simulation framework between the Simulink-Saber power-supply output stage and the Julia thruster slave.

The co-simulation adopts local TCP request-response communication. In the main case, the communication macro-step is consistent with the port recording interval, approximately 1 μs, while the internal time step of HallThruster.jl is 5 ns. Therefore, within one communication macro-step, the power-port voltage is held by zero-order hold, whereas the internal thruster state is advanced continuously with a smaller time step. This interface is an explicitly synchronized sample-and-hold port coupling, and the returned current has at most one communication macro-step of response delay relative to the input voltage. This delay is much smaller than the low-frequency oscillation period considered in this work and therefore does not affect the dominant-frequency identification in millisecond-scale windows.

This encapsulation preserves the traceability of the original HallThruster.jl model and diagnostic quantities while allowing it to serve as a physical load module callable by the power-supply side. The load current is calculated from the internal thruster state under the current discharge boundary. Accordingly, each $V_{cmd}$ value participates in the subsequent generation of electron density, neutral density, potential, axial electric field, electron temperature, and discharge current.

The interface layer transmits only two port variables, $V_{cmd}$ and $I_{out}$, to maintain a clear coupling boundary between the power-supply side and the thruster side. To improve long-term stability, the communication layer performs finite-value checks on the input voltage and returned current, preventing NaN, Inf, or other abnormal values from entering subsequent exchanges. The diagnostic outputs are divided into port quantities and internal quantities. The port time series include t, $V_{cmd}$ and $I_{out}$ and are used for the analysis of mean current, peak-to-peak value, local peaks and troughs, and FFT. The internal field quantities include $n_e$,$n_n$,$\phi$,$E_z$and$T_e$and are used to observe the spatiotemporal evolution of the ionization region, acceleration region, and electron-heating region during low-frequency cycles. Both types of data are saved using the same time base.

## 3. Results

Based on the online co-simulation method described above, this section analyzes the coupled response between the propulsion power-supply output stage and the Hall-thruster load. The port-current response and spectral characteristics of the main co-simulation case are first presented. Then, a fixed-boundary reference, internal-field comparisons, and a resistive-load comparison are used to further demonstrate the influence of online closed-loop coupling on the thruster discharge process and the power-supply output response.

### *3.1 Output Characteristics of the Hall Thruster Under Closed-Loop Feedback*

Figure 2 shows the discharge-current response of the main co-simulation case over 0–10 ms. After the start-up transient, the current does not converge to a constant value but maintains periodic peaks and troughs in the subsequent time interval. The local window of 4–5 ms shows a relatively stable pulsed low-frequency waveform. The FFT dominant peaks in the 1–5 ms and 5–10 ms windows are both located near 11 kHz, indicating that this response is not a short-lived start-up disturbance. Table 1 gives the quantitative metrics in different time windows. In the 1–5 ms window, the mean current is 4.945 A, the peak-to-peak value is 14.222 A, and the dominant frequency is 11.50 kHz. In the 5–10 ms window, the mean current is 4.893 A, the peak-to-peak value is 13.085 A, and the dominant frequency is 11.40 kHz. The mean values, amplitudes, and dominant frequencies in the two later windows remain at the same order, indicating that the main case forms a sustained low-frequency dynamic load.

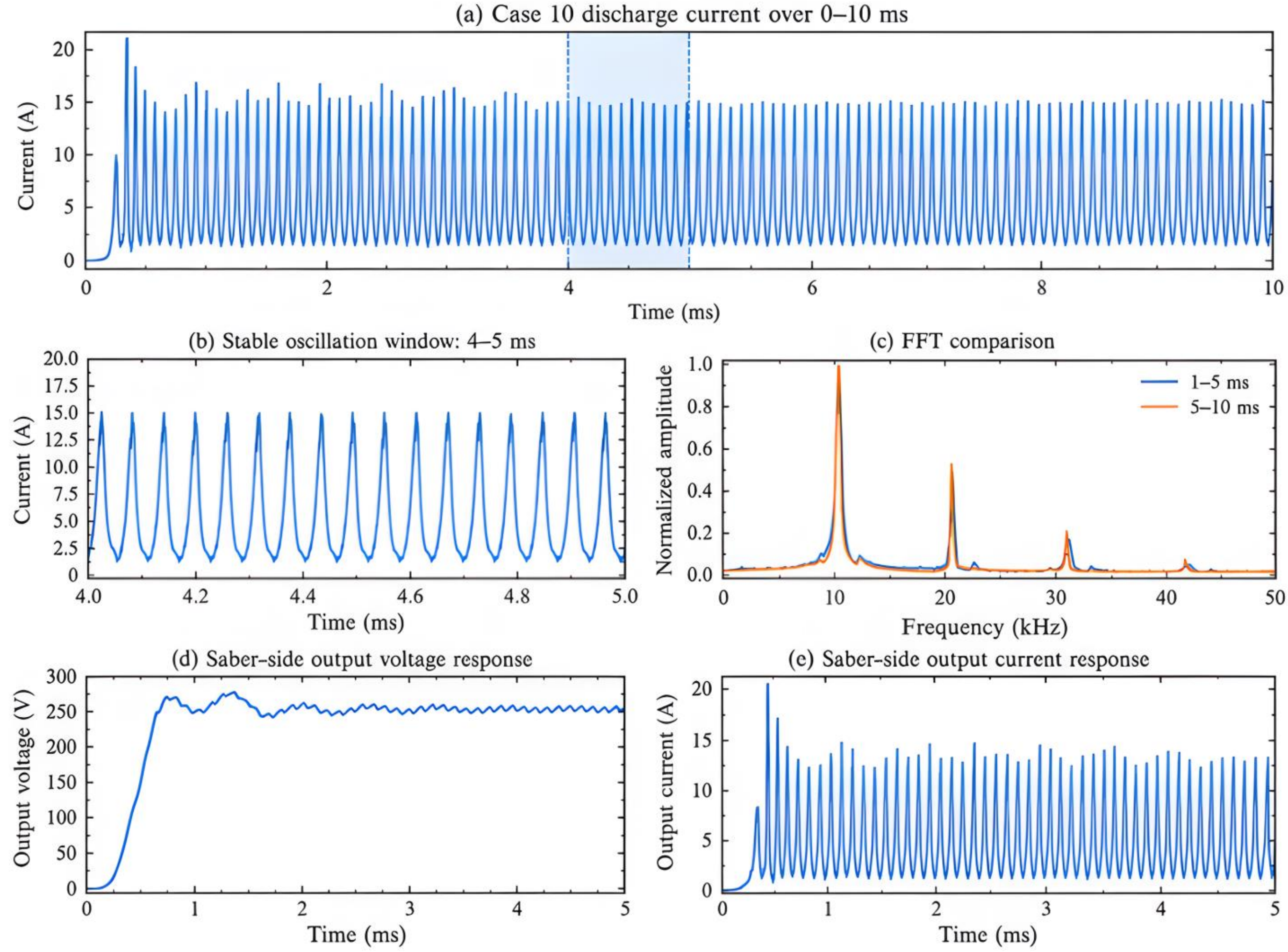

Figure 2. Discharge-current response of the main co-simulation case over 0–10 ms, local oscillation window, and FFT comparison.

Table 2. Discharge-current metrics in different time windows for the main co-simulation case.

| Window | Mean current (A) | Standard deviation (A) | Peak-to-peak value (A) | Dominant frequency (kHz) |
|---|---|---|---|---|
| 1–5 ms | 4.945 | 3.704 | 14.222 | 11.500 |
| 4–5 ms | 4.971 | 3.778 | 13.160 | 11.000 |
| 5–10 ms | 4.893 | 3.747 | 13.085 | 11.398 |

To determine whether the online co-simulation merely reproduces the self-oscillation of the thruster under fixed-voltage simulation, this study compares the co-simulation and fixed-voltage simulation results in the 4.20–4.95 ms window. The reference uses a fixed discharge parameter of 245 V, matched to the effective port voltage in the later stage of the co-simulation. This window contains approximately eight low-frequency cycles and is suitable for comparing peaks, troughs, and phase.

Figure 3 shows that obvious low-frequency periodic peaks exist under both boundary conditions, indicating that this frequency band is consistent with the internal ionization-neutral coupling process of the thruster. Meanwhile, the two current curves do not completely coincide in peak location, local spike shape, or phase, indicating that the online power-supply boundary modulates the discharge cycle. Table 3 shows that the mean currents and peak-to-peak values of the two cases are close but still different. The fixed-voltage simulation demonstrates that the low-frequency oscillation is not generated artificially by the power-supply model, whereas the co-simulation indicates that the dynamic port condition of the output stage has changed the instantaneous response of the thruster as a load.

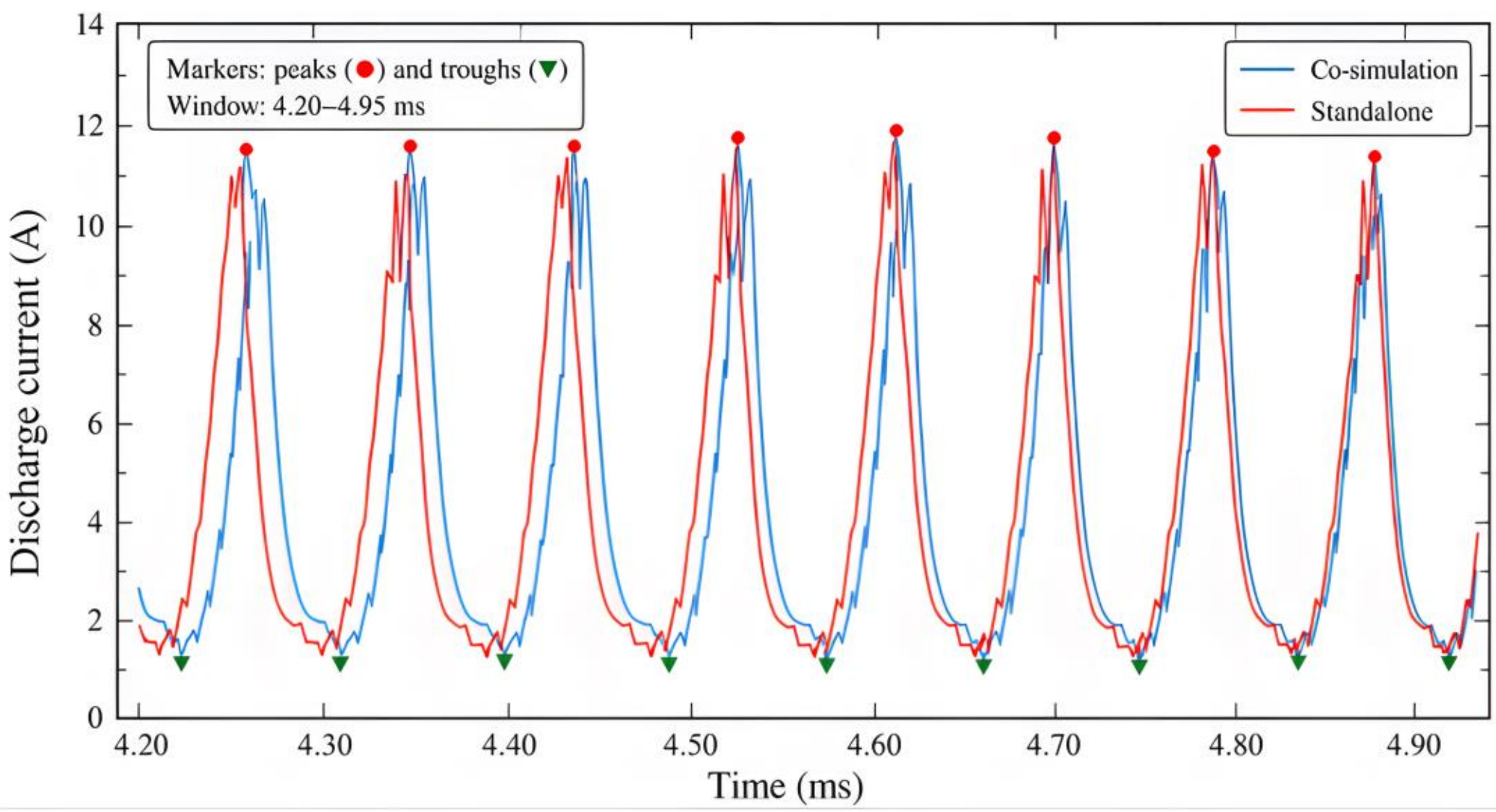

Figure 3. Discharge-current comparison between the co-simulation and fixed-voltage simulation in a stable eight-cycle window.

Table 3. Local discharge-current metrics of the co-simulation and fixed-voltage simulation (4.20–4.95 ms).

| Metric | Online co-simulation | Fixed-voltage simulation |
| --- | --- | --- |
| Mean current (A) | 4.741 | 4.767 |
| Standard deviation (A) | 3.722 | 3.794 |
| Peak-to-peak value (A) | 12.998 | 13.050 |

Figure 4. Comparison of peak/trough time occupancy between the co-simulation and fixed-voltage simulation.

Figure 4 further compares the high-current and low-current stages under the two conditions. The peak stage is defined as the interval in which the current is higher than the window mean, and the trough stage is defined as the interval in which the current is lower than or equal to the window mean. The peak-stage time occupancy is 35.87% for the co-simulation and 34.55% for the fixed-boundary reference, while the corresponding trough-stage occupancies are 64.13% and 65.45%, respectively. Both cases exhibit non-sinusoidal low-frequency oscillations characterized by short peaks and long troughs, but the high-current stage lasts slightly longer in the co-simulation.

Figure 5 gives the FFT comparison between the co-simulation and fixed-voltage reference in the 1–5 ms window. The dominant peak of the online co-simulation is 11.50 kHz, and that of the fixed-boundary reference is 10.99 kHz, with a difference of approximately 0.51 kHz. Combined with the values of 11.309 kHz and 11.067 kHz obtained from the local-peak method, the two cases can be regarded as belonging to the same breathing-mode frequency range, but they do not exhibit exactly the same periodic response. Under fixed-voltage simulation, the cycle is mainly determined by neutral replenishment, enhanced ionization, and depletion. In the co-simulation, the output-port voltage changes slightly with the discharge current, thereby modifying the instantaneous conditions for ionization enhancement and recovery within each cycle.

In addition to the dominant peak, the two spectra also differ at the second harmonic and at higher frequencies. This indicates that the influence of dynamic voltage appears not only as a frequency shift but also as changes in spike sharpness and non-sinusoidal waveform structure. Therefore, the FFT results and local peak/trough statistics jointly show that the main low-frequency oscillation originates from thruster discharge dynamics, while the power-supply parameters change its port-level expression.

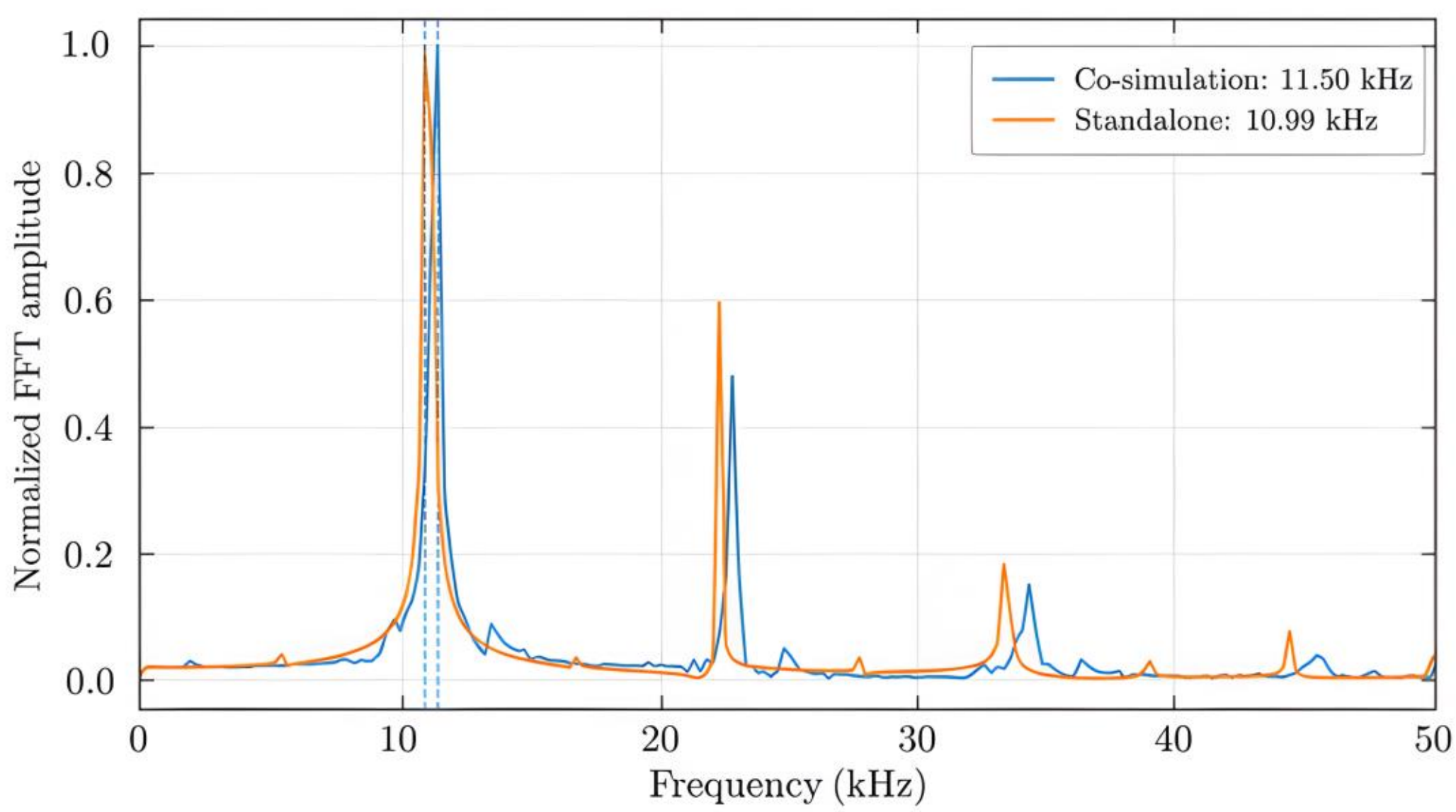

Figure 5. Discharge-current FFT comparison between the co-simulation and fixed-voltage simulation (1–5 ms).

Table 4. FFT dominant-peak comparison (1–5 ms).

| Case | Sampling interval (μs) | FFT dominant frequency (kHz) | Dominant-peak amplitude (A) |
|---|---|---|---|
| Online co-simulation | 1.000 | 11.50 | 3.428 |
| Fixed-boundary reference | 2.500 | 10.99 | 3.385 |

The difference in port current can only indicate a change in the output response; it cannot alone prove that the internal thruster solution is affected by the dynamic boundary. To further verify whether the online coupling enters the internal physical process, this study compares the spatiotemporal distributions of $n_e$, $n_n$, $\phi$, $E_z$ and $T_e$ in a local window. This window covers approximately four low-frequency cycles and can reveal the periodic evolution of the ionization region, acceleration region, and electron-heating region. The quantities include $n_e$, $n_n$, $\phi$, $E_z$ and $T_e$ are directly solved or diagnosed by HallThruster.jl during time advancement and are not transmitted to Saber as port variables. Therefore, these fields provide internal validation independent of the $V_{cmd}$ - $I_{out}$ exchange. If spatiotemporal structural differences appear between the co-simulation and the fixed-voltage simulation, it indicates that the effect of the dynamic port boundary has entered the internal thruster discharge evolution rather than remaining only at the external current-waveform level.

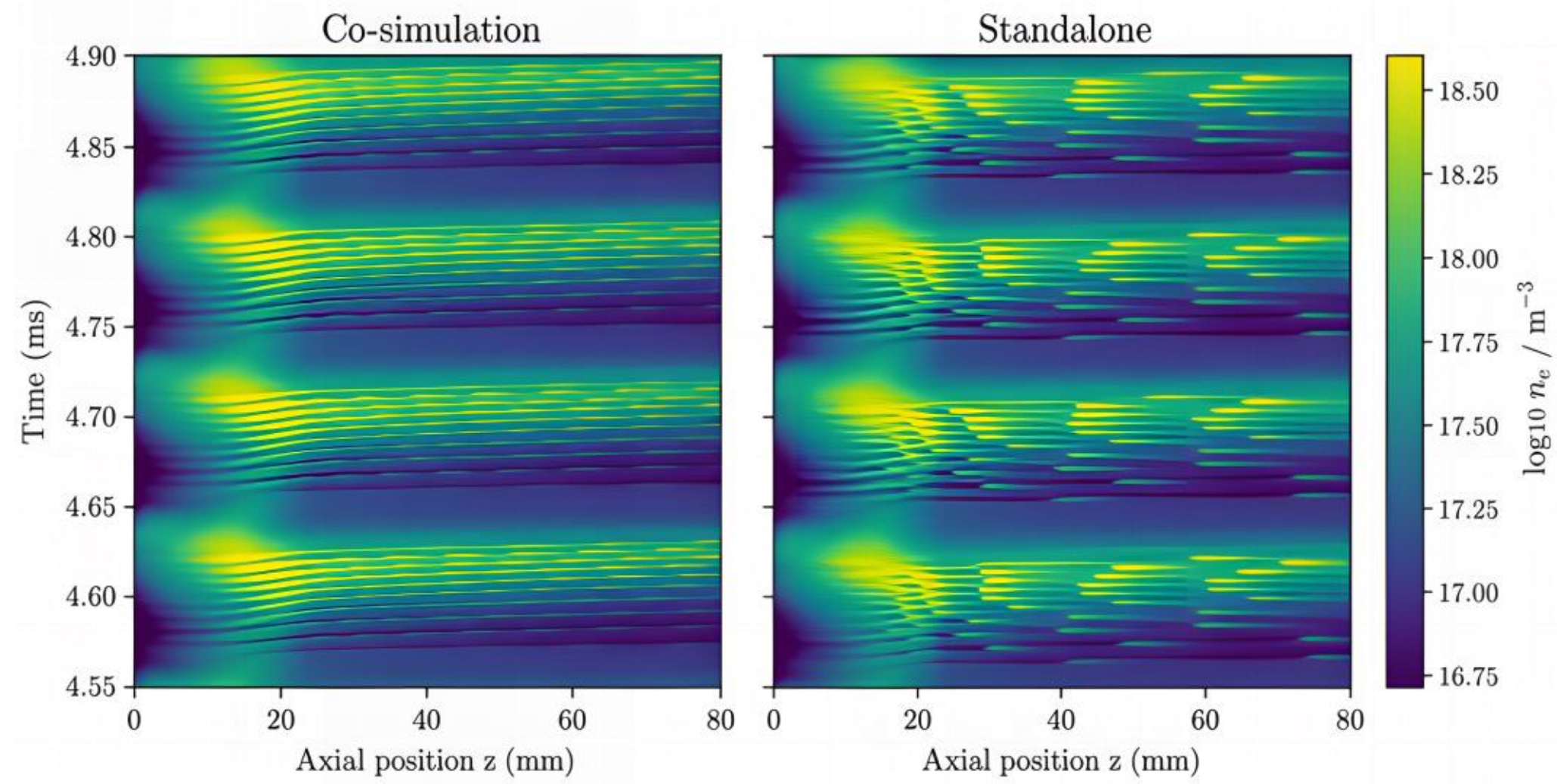

Figure 6. Local comparison of electron density $n_e$ between the co-simulation and fixed-voltage simulation.

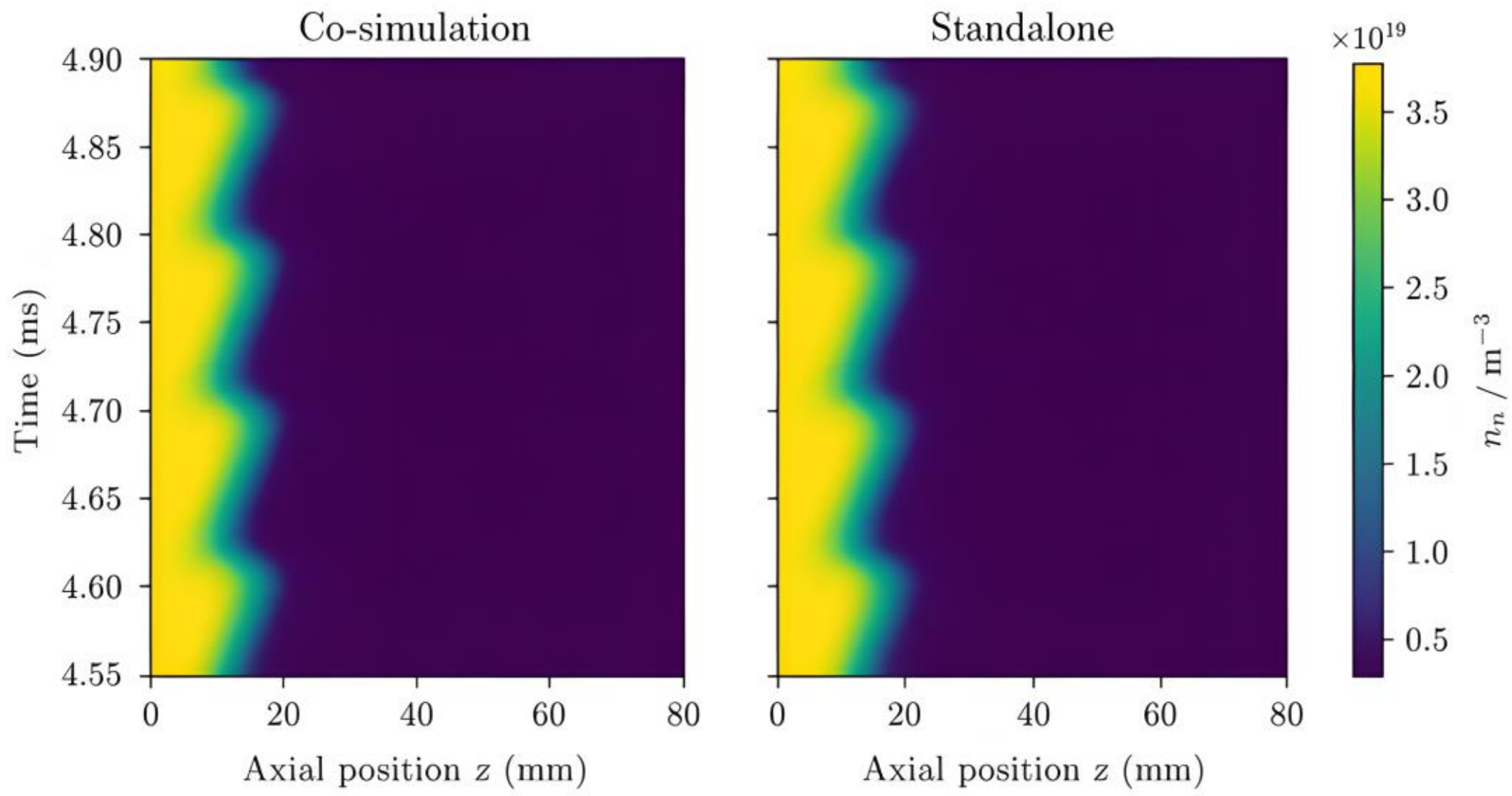

Figure 7. Local comparison of neutral density $n_n$ between the co-simulation and fixed-voltage simulation.

Figures 6 and 7 show the local spatiotemporal distributions of electron density and neutral density, respectively. Under both boundary conditions, repeated enhancement stripes appear over time, and nn exhibits periodic depletion and replenishment near the main ionization region. This correspondence is consistent with the physical picture of the breathing mode, in which neutral replenishment, enhanced ionization, and particle depletion occur alternately. The two results differ in the position of the high-density region, stripe continuity, and depletion boundary, indicating that the dynamic port boundary has affected the local spatiotemporal evolution of the ionization process.

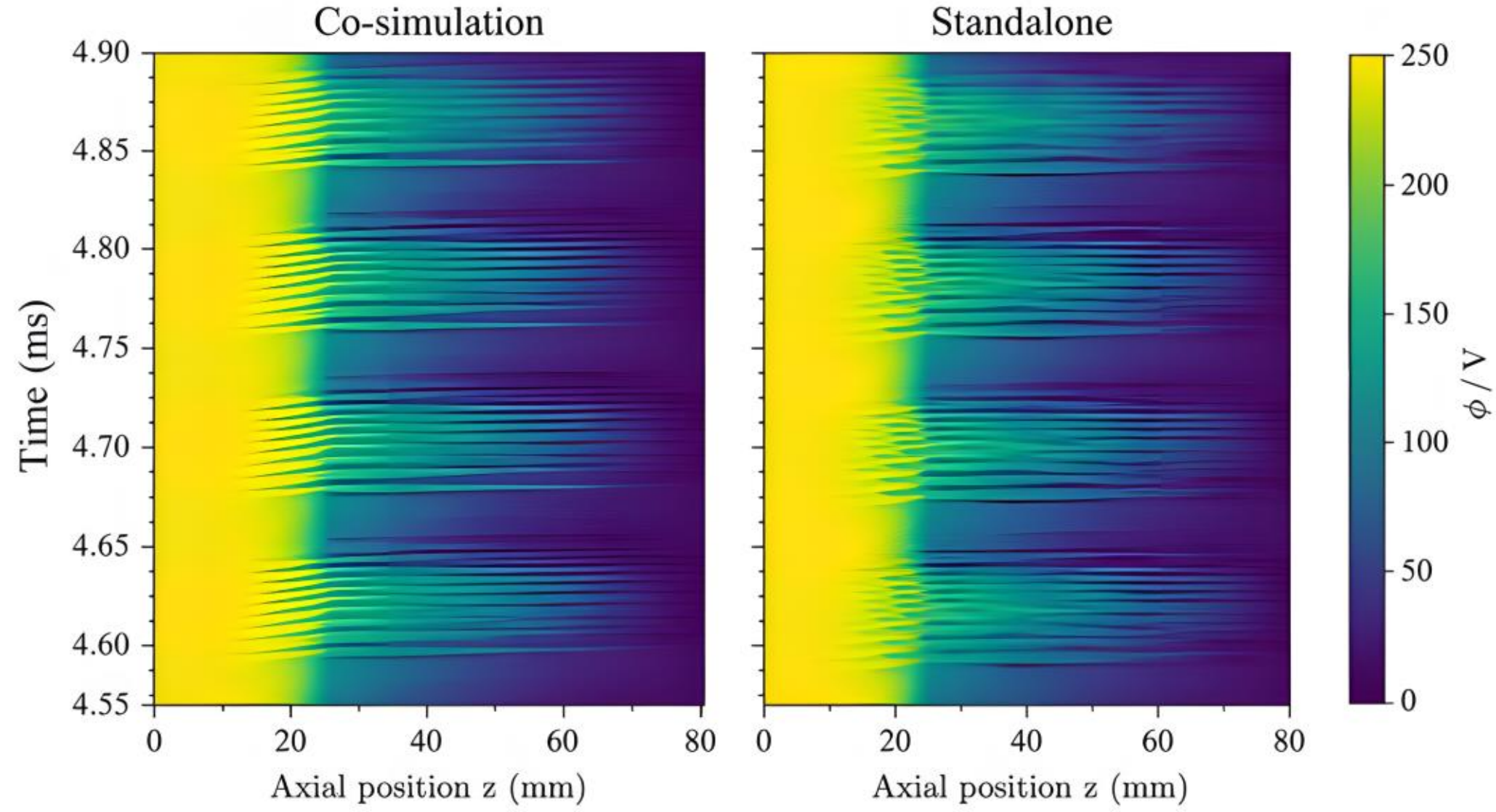

Figure 8. Local comparison of potential $\phi$ between the co-simulation and fixed-voltage simulation.

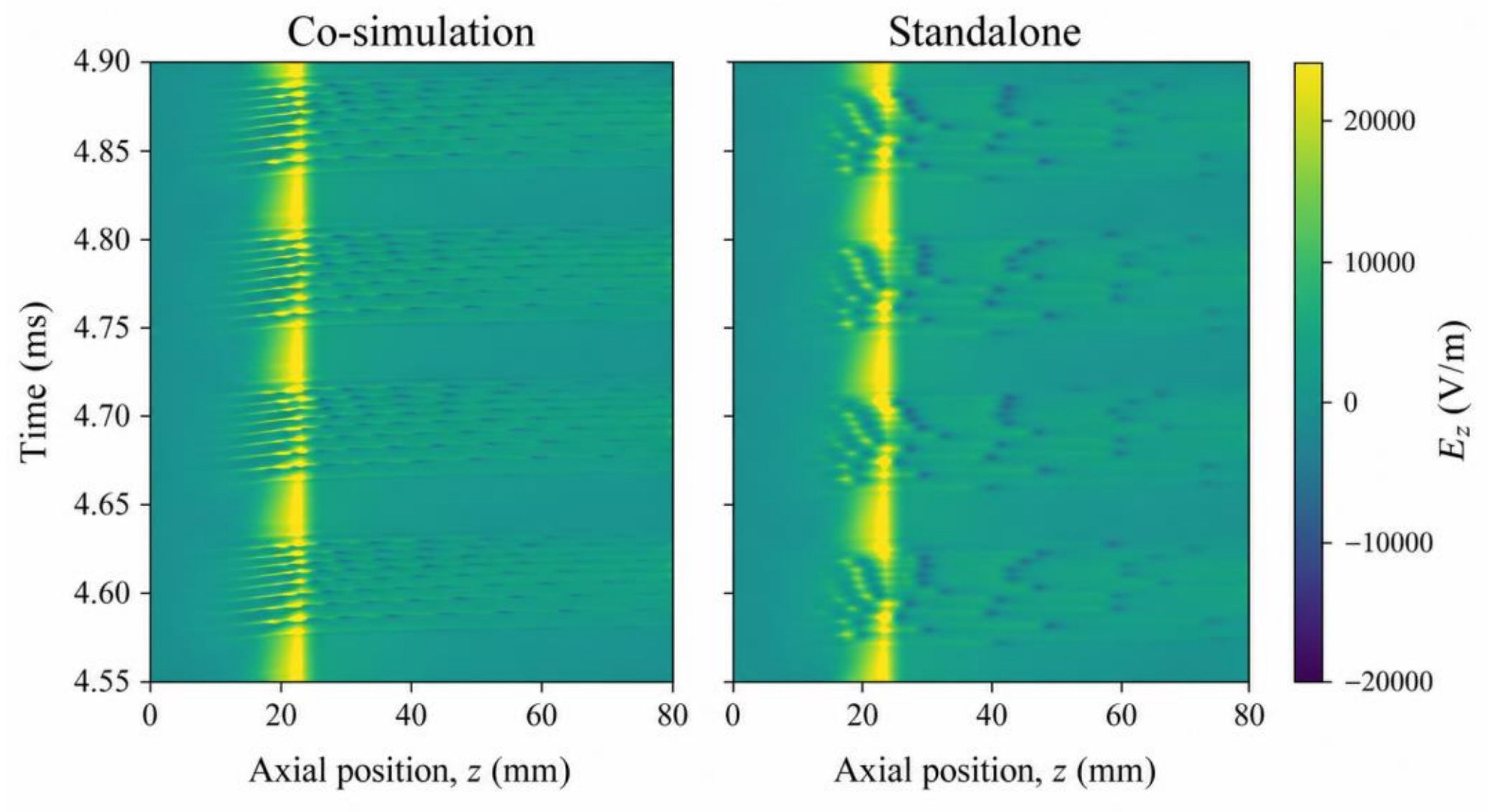

Figure 9. Local comparison of axial electric field $E_z$ between the co-simulation and fixed-voltage simulation.

Figures 8 and 9 show local comparisons of potential and axial electric field. Under both boundary conditions, the main potential-drop region and strong-electric-field region can be observed to repeat with the low-frequency cycle, indicating that the acceleration-region structure corresponds to the discharge-current oscillation. Compared with the fixed-boundary reference, the potential distribution and strong-electric-field region in the online co-simulation differ in local phase, stripe bending, and high-value-region position, indicating that the effect of the dynamic boundary has extended from the port current to the ion-acceleration-related region.

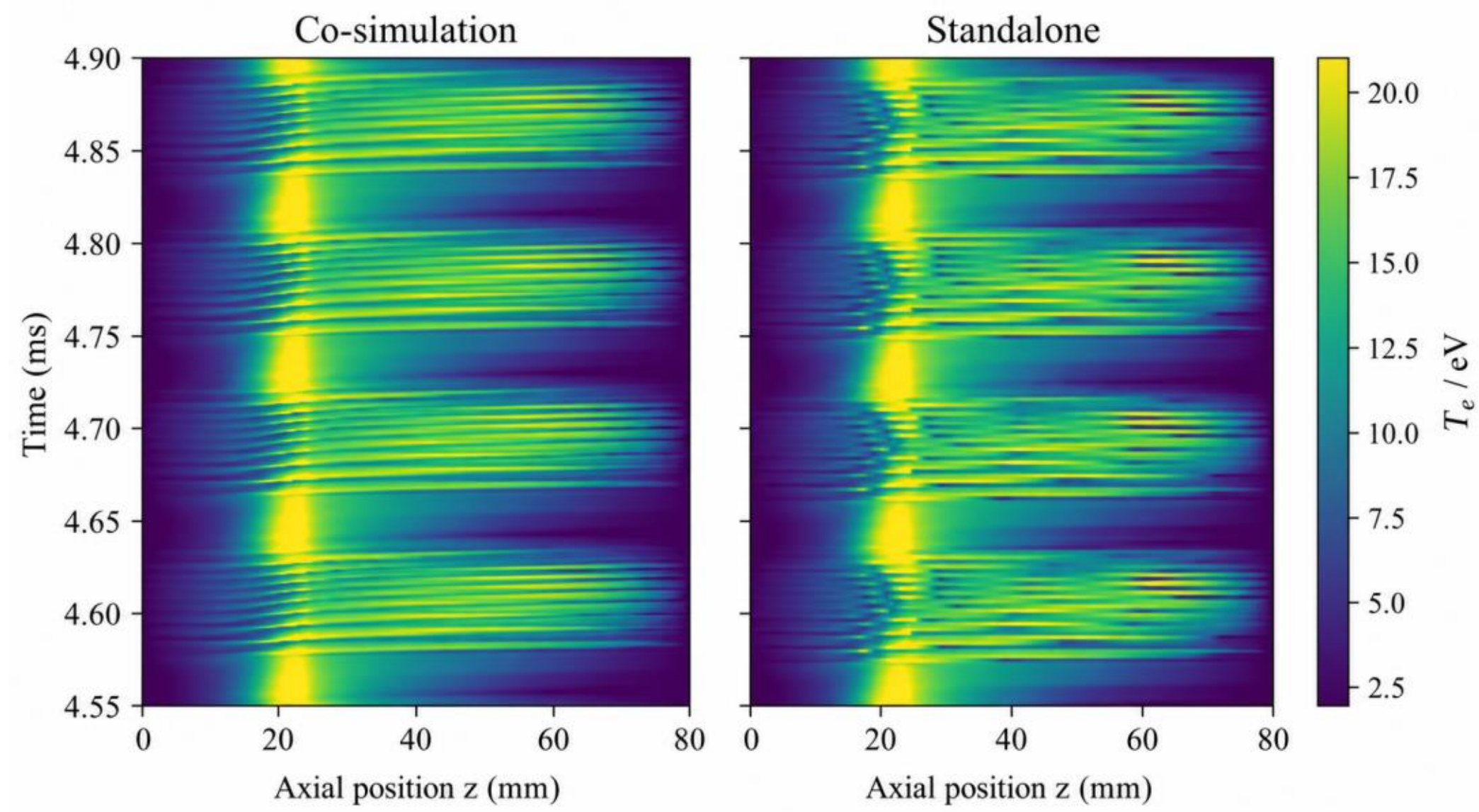

Figure 10. Local comparison of electron temperature $T_e$ between the co-simulation and fixed-voltage simulation.

Figure 10 shows the local spatiotemporal distribution of electron temperature. The high-value region spatially corresponds to the strong-electric-field region and repeats with the low-frequency cycle, indicating that electron heating, ionization enhancement, and acceleration-region adjustment are coupled within the same discharge dynamics. The differences in morphology and local phase between the co-simulation and the fixed-voltage simulation further indicate that the dynamic port boundary can affect the electron-energy state.

Figures 6-10 show that the two calculations maintain similar low-frequency current scales, but the local spatiotemporal evolution of the internal fields differs. Electron density, neutral density, potential, axial electric field, and electron temperature are all solved by HallThruster.jl under the current boundary condition and are not transmitted as port variables. Therefore, these differences demonstrate that the port boundary in the online co-simulation has entered the internal thruster discharge process and has produced observable effects on the plasma structure. The above results verify the effectiveness of the online co-simulation link from three aspects: port response, spectral characteristics, and internal field quantities. The power-supply output stage is no longer only an externally prescribed boundary, and the thruster is no longer a fixed load; instead, they interact within the same time-marching process through the $V_{cmd}$- $I_{out}$ port variables.

### *3.2 Comparison of Power-Supply Response Characteristics Under Resistive and Hall-Thruster Loads*

To demonstrate the dynamic characteristics of a Hall thruster as a propulsion power-supply load, this study compares the port-voltage responses of the same power-supply output stage under resistive-load and Hall-thruster-load conditions. The resistive-load case is used to characterize the basic voltage rise, filtering, and damping characteristics of the output stage under a fixed-load condition. The Hall-thruster-load case is obtained from the online co-simulation, and its port-load characteristics vary in real time with the internal thruster discharge state.

Figure 11 compares the output-voltage responses under the resistive load and the Hall-thruster dynamic load. In the 0–5 ms response, the output voltage under the resistive load rises rapidly during start-up, exhibits an overshoot near 1 ms, and then gradually damps before approaching a steady value. This process mainly reflects the transient response of the power-supply output filter network and energy-storage components under a fixed load. In comparison, under the Hall-thruster load, the output voltage also completes the voltage-rise process, but the later-stage waveform still exhibits sustained periodic fluctuations, and the overall operating-voltage level is lower than that of the resistive-load case.

The enlarged response over 4–5 ms further shows the difference between the two load conditions in the later stage. In this stable window, the output voltage under the resistive load remains approximately around 262 V, with a peak-to-peak value of about 2.21 V, indicating that the power-supply output stage has reached a relatively steady response under the fixed-load condition. In contrast, under the Hall-thruster load, the mean output voltage is approximately 245 V and the peak-to-peak value is about 11.85 V, which is significantly higher than that of the resistive-load case. This result indicates that the Hall-thruster dynamic load introduces stronger periodic modulation into the power-supply output port.

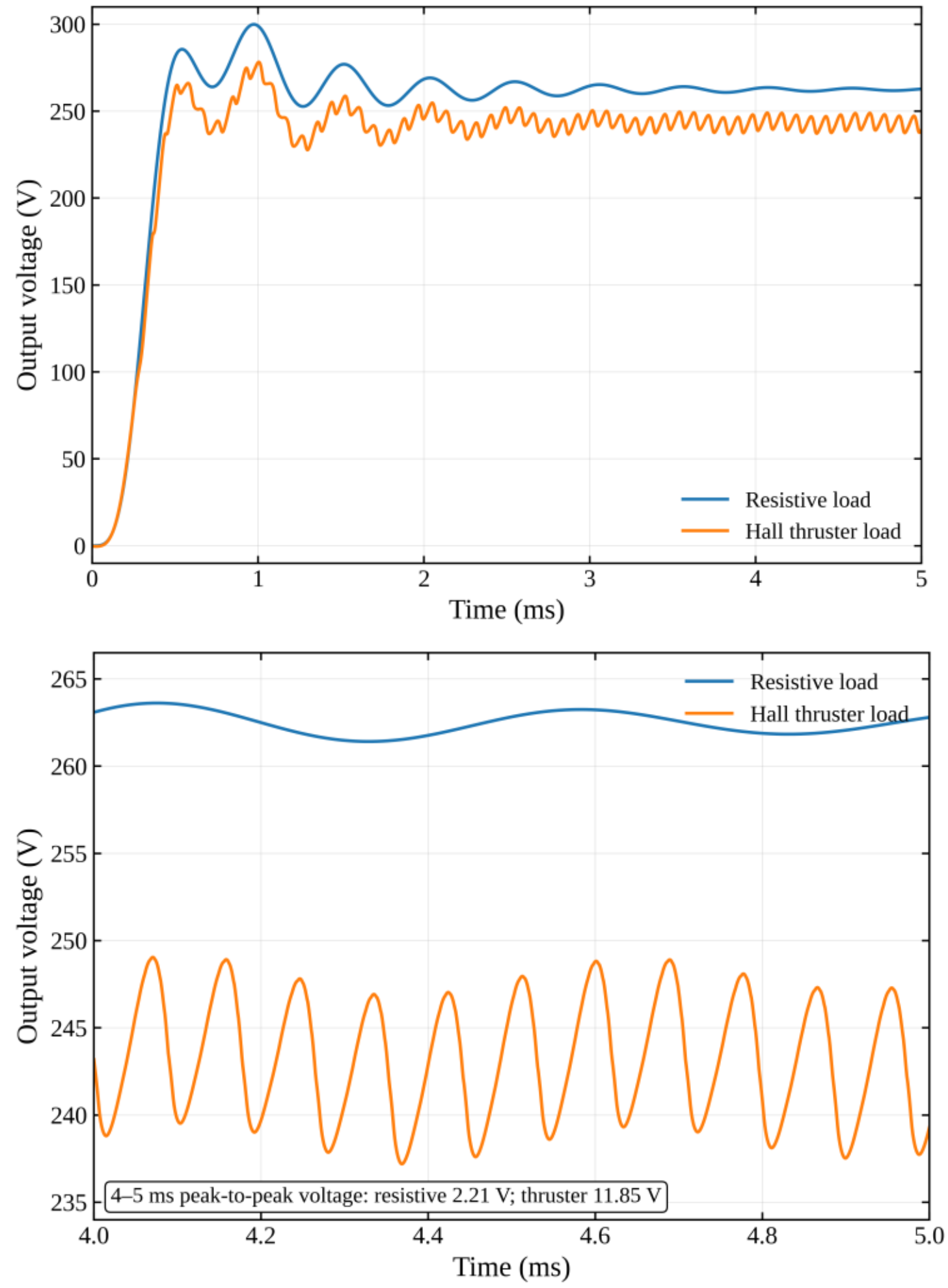

Figure 11. Global and local output-voltage responses under resistive and Hall-thruster loads.

The difference between the two load responses originates from the different load mechanisms. The port relationship of the resistive load is mainly determined by a fixed resistance, and the output-voltage variation is primarily controlled by the parameters of the power-supply output stage itself. In contrast, the Hall-thruster load is jointly determined by internal discharge processes such as neutral transport, ionization, electron transport, potential distribution, and acceleration-region evolution. The periodic variation of the internal thruster state changes the equivalent load characteristics and feeds back to the power-supply side through the port current, causing sustained ripple in the output voltage.

Therefore, resistive-load simulation is suitable for checking the basic voltage-rise capability and steady-state output capability of the power-supply output stage, but it cannot represent the dynamic load environment of a Hall thruster during actual operation. If only a fixed resistive load is used in PPU output-stage analysis, the influence of the thruster discharge process on port-voltage ripple and the output filter network may be underestimated. In contrast, the online co-simulation based on HallThruster.jl-Saber/Simulink can introduce physical load characteristics that vary with the discharge state into the power-supply simulation, providing more reasonable simulation inputs for subsequent output-ripple evaluation and component-stress analysis.

## 4. Discussion

To address the difficulty of real-time coupling of Hall-thruster loads in propulsion power-supply output-stage simulation, this paper establishes a HallThruster.jl-Simulink-Saber port-level online co-simulation method. In this method, the output-port voltage from Simulink-Saber is used as the instantaneous discharge boundary of HallThruster.jl, and the discharge current Iout calculated from the internal thruster state is returned to the power-supply output stage. In this way, synchronized closed-loop coupling between the propulsion power-supply output stage and the one-dimensional thruster discharge model is achieved.

The main co-simulation case shows that, after the start-up transient, the discharge current forms a sustained low-frequency periodic oscillation. In the two later windows of 1-5 ms and 5-10 ms, the dominant discharge-current frequencies are approximately 11.50 kHz and 11.40 kHz, and the peak-to-peak values are approximately 14.22 A and 13.09 A, respectively. This indicates that the load observed by the power-supply side is not a short-term start-up transient, but a dynamic load response continuously generated by the internal thruster discharge process. Compared with the fixed-boundary reference, the online co-simulation maintains similar mean-current levels and dominant-frequency ranges, but differs in peak positions, local waveform, high/low-current duration, and harmonic structure, indicating that the port boundary of the power-supply output stage modulates the thruster discharge cycle.

The internal-field comparison further shows that although ne,nn,ϕ,$E_z$ and Te are not transmitted as port variables, observable differences in local spatiotemporal structures still appear between the online co-simulation and the fixed-boundary reference. This result demonstrates that the influence of the dynamic port boundary has entered the internal HallThruster.jl discharge solution rather than remaining only at the external current-waveform level. The resistive-load comparison also shows that a fixed load can only reflect the basic voltage-rise and steady-state output capabilities of the power-supply output stage and cannot describe the sustained effect of the Hall-thruster discharge process on port-voltage ripple.

It should be noted that the Saber-side model in this study is mainly used to represent the port boundary of the propulsion power-supply output stage, and a complete PPU control, protection, and multi-output power-supply system has not yet been considered. Therefore, the results of this paper are primarily used to verify the port-level online coupling method between the physical thruster model and the power-supply output stage. In future work, a more complete PPU topology, control strategy, and component parasitic parameters can be introduced into this framework to conduct quantitative analyses of output ripple, dynamic-load matching, and key-component stress.

**Author Contributions:** Conceptualization, Z.F., Y.Z. and L.W.; methodology, Z.F. and Y.Z.; software, Z.F.; validation, Z.F.; formal analysis, Z.F.; investigation, Z.F.; resources, L.W.; data curation, Z.F.; writing—original draft preparation, Z.F.; writing—review and editing, Y.Z., J.L., Y.T., L.S., S.W. and L.W.; visualization, Z.F.; supervision, Y.Z. and L.W.; project administration, L.W. All authors have read and agreed to the published version of the manuscript.

Funding: research was funded by the Key Research and Development Program of Heilongjiang Province, grant number 2024ZX05B07, and the Advanced Space Propulsion

Laboratory of BICE and Beijing Engineering Research Center of Efficient and Green Aerospace Propulsion Technology, grant number LabASP-2023-02.

Data Availability Statement: The data presented in this study are available from the corresponding author upon reasonable request.

Acknowledgments: The authors would like to thank the members of the research group for their helpful discussions and technical support during the development of the HallThruster.jl–Saber/Simulink co-simulation platform.

Conflicts of Interest: The authors declare no conflicts of interest.

## References

1. D. Katz Franco, B. Shoor, A. Davidson, R. Zimmerman, L. Appel, V. Rinski, D. Lev, and J. Herscovitz, "On-orbit mission overview of the low power Hall thruster propulsion system aboard Venμs satellite," Journal of Electric Propulsion, vol. 3, Art. no. 15, 2024.
2. The Aerospace Corporation, "2023 Small Satellite Propulsion Technologies Survey," Tech. Rep., 2023.
3. D. M. Goebel and I. Katz, Fundamentals of Electric Propulsion: Ion and Hall Thrusters. Hoboken, NJ, USA: Wiley, 2008.
4. O. Petrenko, O. Alekseenko, and V. Tolmachov, "Power processing and control unit of the electric propulsion system," Journal of Rocket-Space Technology, vol. 32, no. 4, pp. 15–22, 2023.
5. M. Fang, D. Zhang, and X. Qi, "A novel power processing unit (PPU) system architecture based on HFAC bus for electric propulsion spacecraft," IEEE Journal of Emerging and Selected Topics in Power Electronics, vol. 10, no. 5, pp. 5381–5391, 2022.
6. C. R. Rhodes, G. F. Benavides, and L. R. Piñero, "Sub-kW class Hall-effect thruster power processing unit for wide output range applications," in Proc. 38th International Electric Propulsion Conference, Toulouse, France, 2024, IEPC-2024-331.
7. E. Y. Choueiri, "Plasma oscillations in Hall thrusters," Physics of Plasmas, vol. 8, no. 4, pp. 1411–1426, 2001.
8. J.-P. Boeuf and L. Garrigues, "Low frequency oscillations in a stationary plasma thruster," Journal of Applied Physics, vol. 84, no. 7, pp. 3541–3554, 1998.
9. T. Lafleur, P. Chabert, and A. Bourdon, "The origin of the breathing mode in Hall thrusters and its stabilization," Journal of Applied Physics, vol. 130, no. 5, Art. no. 053305, 2021.
10. E. T. Dale and B. A. Jorns, "Experimental characterization of Hall thruster breathing mode dynamics," Journal of Applied Physics, vol. 130, no. 13, Art. no. 133302, 2021.
11. O. Chapurin, A. Smolyakov, G. J. M. Hagelaar, and Y. Raitses, "On the mechanism of ionization oscillations in Hall thrusters," Journal of Applied Physics, vol. 129, no. 23, Art. no. 233307, 2021.
12. O. Chapurin, A. Smolyakov, G. J. M. Hagelaar, J.-P. Boeuf, and Y. Raitses, "Fluid and hybrid simulations of the ionization instabilities in Hall thrusters," Journal of Applied Physics, vol. 132, no. 5, Art. no. 053301, 2022.
13. F. Petronio, A. Alvarez Laguna, A. Bourdon, and P. Chabert, "Study of the breathing mode development in Hall thrusters using hybrid simulations," Journal of Applied Physics, vol. 135, no. 7, Art. no. 073301, 2024.
14. L. Leporini, V. Giannetti, M. M. Saravia, F. Califano, S. Camarri, and T. Andreussi, "On the onset of breathing mode in Hall thrusters and the role of electron mobility fluctuations," Frontiers in Physics, vol. 10, Art. no. 951960, 2022.
15. L. Leporini, V. Giannetti, M. M. Saravia, F. Califano, S. Camarri, and T. Andreussi, "An unstable 0D model of ionization oscillations in Hall thrusters," Frontiers in Physics, vol. 10, Art. no. 1097813, 2023.
16. J. A. Gottfried, S. Antozzi, J. Stienike, S. J. Thompson, J. D. Williams, and A. P. Yalin, "Temporally resolved relative krypton neutral density during breathing mode of a Hall effect thruster recorded by TALIF," Journal of Electric Propulsion, vol. 3, Art. no. 9, 2024.
17. L. Brieda, J. Koo, and M. Scharfe, "Influence of a power supply model on simulated Hall thruster discharge voltage oscillations," AIP Advances, vol. 9, no. 2, Art. no. 025320, 2019.
18. J. Simmonds, Y. Raitses, A. Smolyakov, and O. Chapurin, "Studies of a modulated Hall thruster," Plasma Sources Science and Technology, vol. 30, no. 5, Art. no. 055011, 2021.
19. D. R. Jovel, J. D. Cabrera, and M. L. R. Walker, "Hall effect thruster impedance characterization in ground-based vacuum test facilities," Journal of Electric Propulsion, vol. 3, Art. no. 31, 2024.
20. B. A. Jorns, E. Dale, and R. R. Hofer, "Mode transitions in a magnetically shielded Hall thruster. I. Experimentally informed model," Journal of Applied Physics, vol. 136, no. 5, Art. no. 053301, 2024.

21. T. Marks, P. Schedler, and B. Jorns, "HallThruster.jl: a Julia package for 1D Hall thruster discharge simulation," Journal of Open Source Software, vol. 8, no. 86, Art. no. 4672, 2023.
22. The MathWorks, Inc., "Simulink," version 8.6 (R2015b), Natick, MA, USA, 2015.
23. Synopsys, Inc., "Synopsys Saber," version 2016, Mountain View, CA, USA, 2016.